\documentclass{ws-ijmpe}

\begin{document}

\markboth{Signal, Bissey and Leinweber}{Di-quarks and Tri-quarks on the Lattice}

\catchline{}{}{}{}{}
%%%%%%%%%%%%%%%%%%%%%%%%%%%%%%%%%%%%%%%%%%%%%%%%%%%%%%%%%%%%%%%%%%%%

\title{DI-QUARKS AND TRI-QUARKS ON THE LATTICE
}

\author{A. I. SIGNAL\footnote{a.i.signal@massey.ac.nz} 
and F. R. P. BISSEY\footnote{f.r.bissey@massey.ac.nz}}

\address{Institute of Fundamental Sciences, Massey University, Private Bag 11-222,\\
Palmerston North, New Zealand }

\author{D. B. LEINWEBER\footnote{dleinweb@physics.adelaide.edu.au}}

\address{Centre for the Subatomic Structure of Matter and Department of Physics,\\ 
University of Adelaide, Adelaide, SA 5005, Australia }

\maketitle

\begin{history}
\received{(received date)}
\revised{(revised date)}
%\accepted{(Day Month Year)}
%\comby{(xxxxxxxxxx)}
\end{history}

\begin{abstract}
The distribution of gluon fields in hadrons is of fundamental interest in QCD.  
Using lattice QCD we have observed the formation of gluon flux tubes within tri-quark 
(baryon) systems for a wide variety of spatial distributions of the color sources. 
In particular we have investigated configurations where two of the quarks are 
close together and the third quark is some distance away, which approximates a quark 
plus diquark string.
We find that the string tension of the quark - diquark string is the same as that of the 
quark - antiquark string on the same lattice.
We also compare the longitudinal and transverse profiles of the gluon flux tubes for both 
sets of strings, and find them to be of similar radii and to have similar vacuum suppression.
\end{abstract}

\section{Introduction}

Recently there has been renewed interest in studying the distribution
of quark and gluon fields in the three-quark static-baryon system.
While the earliest studies were inconclusive\cite{Flower:1986ru}, improved
computing resources and analysis techniques now make it possible to
study this system in a quantitative manner\cite{Takahashi:2002bw,Alexandrou:2001ip,deForcrand:2005vv}.
In particular, it is possible to directly compute the gluon field
distribution\cite{Ichie:2002mi,Okiharu:2003vt,Bissey:2005sk,Bissey:2007tq} using
lattice QCD techniques similar to those pioneered in mesonic
static-quark systems\cite{Sommer:1987uz,Bali:1994de,Haymaker:1994fm}.
We first investigate the formation of flux tubes, where vacuum gluon field fluctuations are 
suppressed, in systems where the three quarks are approximately equidistant. 
In order to compare baryonic and mesonic flux tubes, we have further investigated 
three-quark static baryons where two of the quarks are close together and the third is some 
distance away

\section{Baryon Flux tubes on the Lattice}

\begin{figure}
\centering\includegraphics[height=3cm,clip=true]{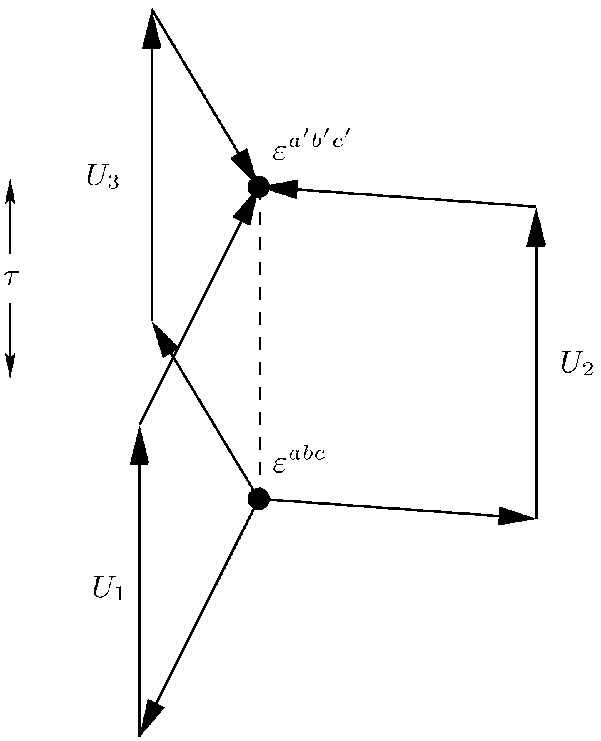}
\caption{Gauge-link paths or ``staples,'' $U_1$, $U_2$ and $U_3$,
  forming a three-quark Wilson loop with the quarks located at $\vec
  r_1$, $\vec r_2$ and $\vec r_3$.  $\varepsilon^{abc}$ and
  $\varepsilon^{a'b'c'}$ denote colour anti-symmetrisation at the source
  and sink respectively, while $\tau$ indicates evolution of the
  three-quark system in Euclidean time.}
\label{staple}
\end{figure}

In order to study flux-tubes on the lattice, we begin with the standard approach of connecting 
static quark  propagators with spatial-link paths in a gauge invariant manner.  
We use APE-smeared spatial-link paths 
(with $\alpha = 0.7$ and 30 smearing steps in all cases)
to propagate the quarks from a common origin to their
spatial positions as illustrated in Fig.~\ref{staple}.  
The static quark propagators are constructed from time directed link products at fixed spatial 
coordinate, $\prod_i U_t(\vec x, t_i)$, using the untouched `thin'' links of the gauge configuration.
Finally smeared-link spatial paths propagate the quarks back to the common
spatial origin.
The three-quark Wilson loop is thus defined as:
\begin{equation}
W_{3Q}=\frac{1}{3!}\varepsilon^{abc}\varepsilon^{a'b'c'} \, U_1^{aa'} \,
U_2^{bb'} \, U_3^{cc'},
\end{equation}  
where $U_j$ is a staple made of path-ordered link variables as shown in 
Fig.~\ref{staple}.
We consider Y and T shaped paths in the $x-y$ plane, where the three quarks are 
created at the origin, then are propagated to positions $\vec r_1$, $\vec r_2$
and $\vec r_3$, which approximate an equilateral triangle, before being propagated through
time and finally back to a sink at the same spatial location as the source. 
We investigate seven such triangles, with average inter-quark distance varying from 
0.27 fm to 1.7 fm.

The gluon-field fluctuations are characterised by the gauge-invariant action 
density $S(\vec y, t)$.  
We calculate the action density using the highly-improved ${\cal O}(a^4)$ 
three-loop improved lattice field-strength tensor \cite{Bilson-Thompson:2002jk} on
four-sweep APE-smeared gauge links.  

For a Wilson loop of Euclidean time extent $\tau$ we evaluate the correlation function
\begin{equation}
C(\vec y; \vec r_1, \vec r_2, \vec r_3; \tau) = 
\frac{
\bigl\langle W_{3Q}(\vec r_1, \vec r_2, \vec r_3; \tau) \,
             S(\vec y, \tau/2) \bigr\rangle }
{
\bigl\langle  W_{3Q}(\vec r_1, \vec r_2, \vec r_3; \tau) \bigr\rangle \,
\bigl\langle S(\vec y, \tau/2) \bigr\rangle
}
  \, ,
\label{correl}
\end{equation} 
where $\langle \cdots \rangle$ denotes averaging over configurations
and lattice symmetries.
This correlates the quark positions with the gauge-field action in a gauge invariant manner, 
and has the advantage of being positive definite, eliminating
any sign ambiguity on whether vacuum field fluctuations are enhanced
or suppressed in the presence of static quarks.  
For $\vec y$ well away from the quark positions, there are no correlations and $C \to 1$. 
We find that $C$ is generally less than 1, signaling the expulsion of vacuum fluctuations
from the interior of heavy-quark hadrons.

We first consider 200 quenched QCD gauge-field configurations created with the 
${\cal O}(a^2)$-mean-field improved Luscher-Weisz plaquette plus rectangle gauge 
action\cite{Luscher:1984xn} on $16^3\times 32$ lattices at $\beta =4.60$.
The long dimension is taken as being the $x$ direction making
the spatial volume $16^2\times 32$.
Using a physical string tension of $\sigma = (0.440\ \text{GeV})^2 =
0.981\ \text{GeV}/\text{fm}$, the $Q\bar{Q}$ potential sets the lattice
spacing to $a=0.123(2)$fm.  

We use lattice symmetries to improve the signal to noise ratio of our measurements. 
These include translational invariance (any point on the lattice can be taken as the origin), 
reflection in the $x$ plane and $90^\circ$ rotational symmetry about the $x$-axis.
The advantage of this approach is that we do not have to perform any gauge fixing to find a 
signal in the flux distributions. 

\begin{figure}
\centering\includegraphics[height=4cm,clip=true]{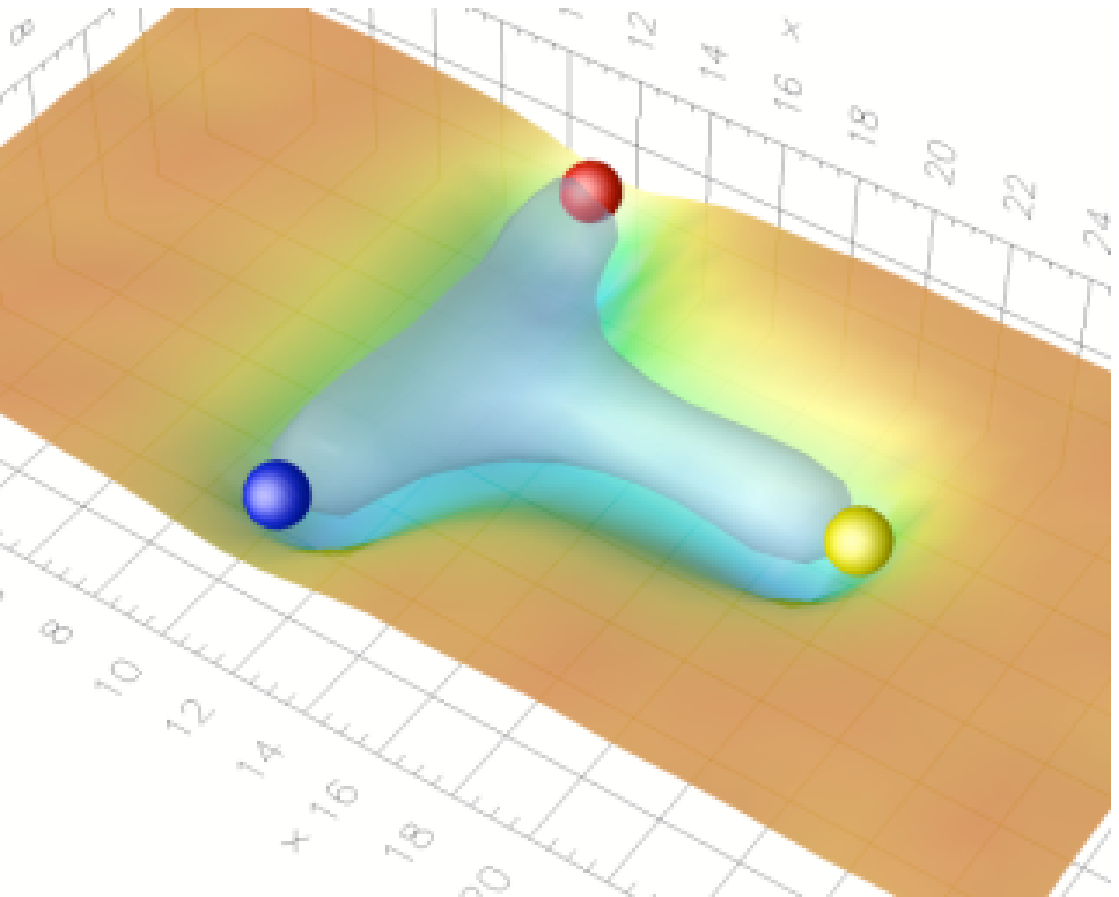}
\centering\includegraphics[height=4cm,clip=true]{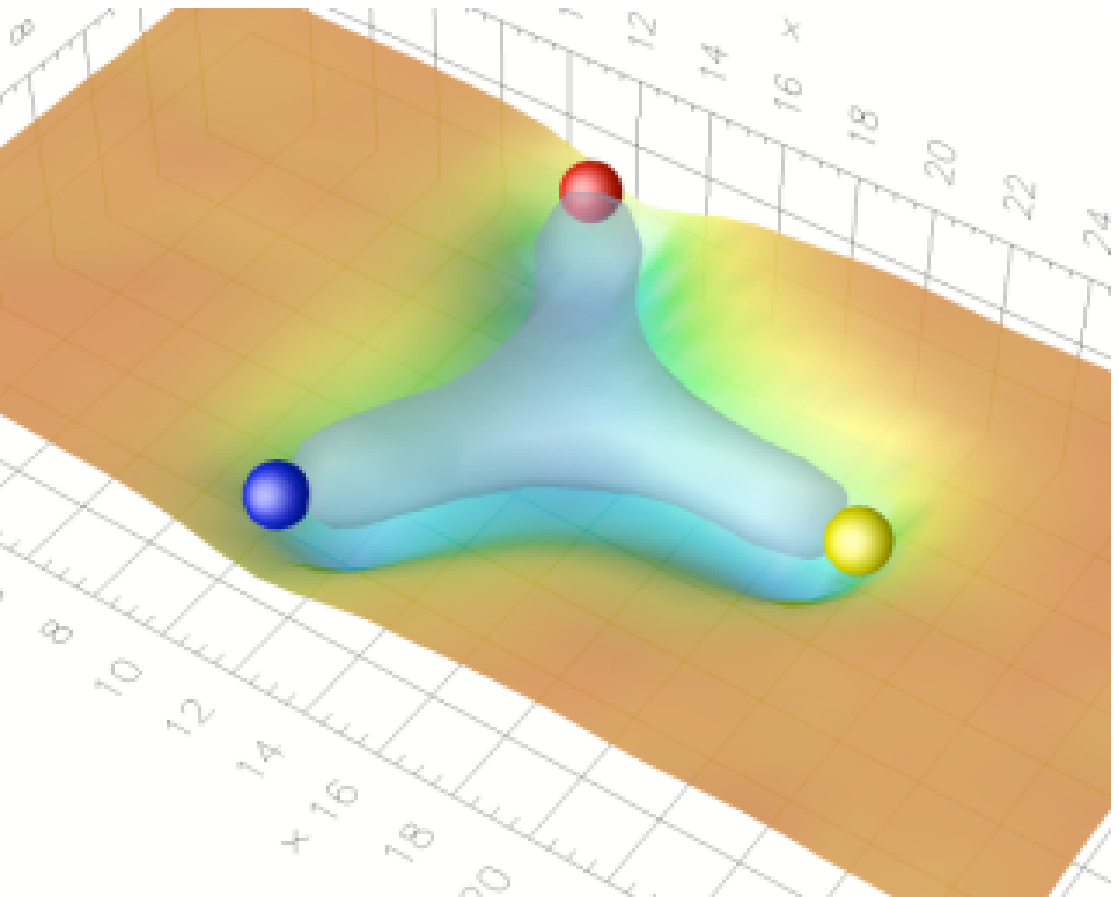}
\caption{Expulsion of gluon-field fluctuations from the region of
  static quark sources illustrated by the spheres.  
  An isosurface of $C(\vec{y})$ is illustrated by the translucent surfaces.  
  The surface plots (or rubber sheets) describes the values of $C(\vec{y})$
  for $\vec y$ in the quark plane, $(y_1, y_2, 0)$.  
  Results are for 30-sweep smeared T-shape (left) and Y-shape (right) sources 
  with the largest quark separation considered.}
\label{Ttube7-s30}
\end{figure}

In Fig.~\ref{Ttube7-s30} we show examples of the expulsion of vacuum 
fluctuations and the formation of flux-tubes for our largest T and Y shaped configurations.

A detailed analysis of these flux tubes is given elsewhere\cite{Bissey:2005sk,Bissey:2007tq}.
The major conclusions can be summarised. 
We do not see any evidence for $\Delta$ - shaped (empty triangle) flux tubes. 
At large quark separations we do see Y shaped flux tubes, even from initial T shapes, 
which relax towards a Y shape.
We observe a potential which is a linear function of the minimum length of string needed to 
connect the quarks to the Fermat point, and the extracted string tension 
($\sigma = 0.97(1)$ GeV fm$^{-1}$) is in good agreement with the quark - antiquark string tension.

\section{Diquarks on the Lattice}

There has recently been a renewal of interest in the properties of diquarks in hadronic systems, as
they may play an important role in the existence of exotic states, such as the putative $\Theta^{+}$, 
or in explaining the scarcity of such exotics\cite{JaffeWilczek}. 
In QCD, two quarks close together, a diquark, can transform either according to the conjugate 
representation $(\bar{3})$ or the sextet $(6)$ representation of $SU(3)$. 
The color hyperfine interaction then leads to attraction in the spin singlet, scalar diquark 
channel, while the spin triplet, axial vector diquark is disfavoured. 
Hence diquarks should have positive parity and belong to the color $\bar{3}$ representation, and 
so have many properties similar to an antiquark. 
In lattice QCD this should lead to the formation of quark - diquark flux tubes with similar physical 
characteristics to those of quark - antiquark flux tubes. 
In particular we would expect the long range linear part of the quark - diquark potential to have the 
same slope as that of the quark - antiquark potential, corresponding to the flux tubes having the 
same energy density, and we would expect the flux tubes to have similar transverse size. 

The three-quark configurations we use to approximate a quark - diquark string are T-shapes, 
with the origin at the junction of the T. Two quarks are positioned one lattice step in opposite 
directions from the origin (approximating the diquark), and the third is placed from 1 to 12 
lattice steps in an orthogonal direction.
For this part of our  work we have used 300 quenched QCD gauge field configurations 
created with the same action as previously.
Two hundred of these configurations  were at at $\beta =4.60$ (as in the previous section) and 
one hundred at $\beta = 4.80$, to investigate the use of a finer lattice.
These values of $\beta$ give lattice spacings $a$ of $0.123$ fm and $0.093$ fm respectively.

\begin{figure}
\centering\includegraphics[height=5cm,clip=true]{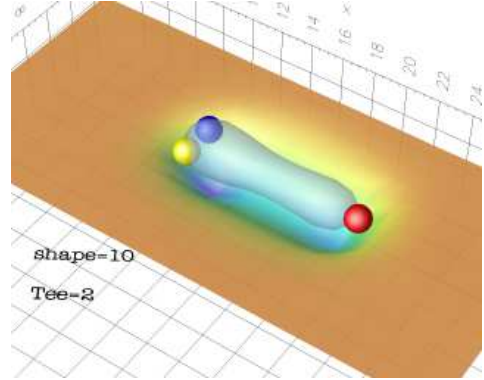}
\caption{Expulsion of gluon-field fluctuations from the region of
  static quark sources illustrated by the spheres.  An isosurface of
  $C(\vec{y})$ is illustrated by the translucent surface.  A
  surface plot (or rubber sheet) describes the values of $C(\vec{y})$
  for $\vec y$ in the quark plane, $(y_1, y_2, 0)$.}
\label{T-tube10}
\end{figure}

In Fig.~\ref{T-tube10} we show an example of the expulsion of vacuum 
fluctuations and the formation of flux-tubes for our  quark - diquark configurations.

\begin{figure}
\centering\includegraphics[width=4cm,clip=true]{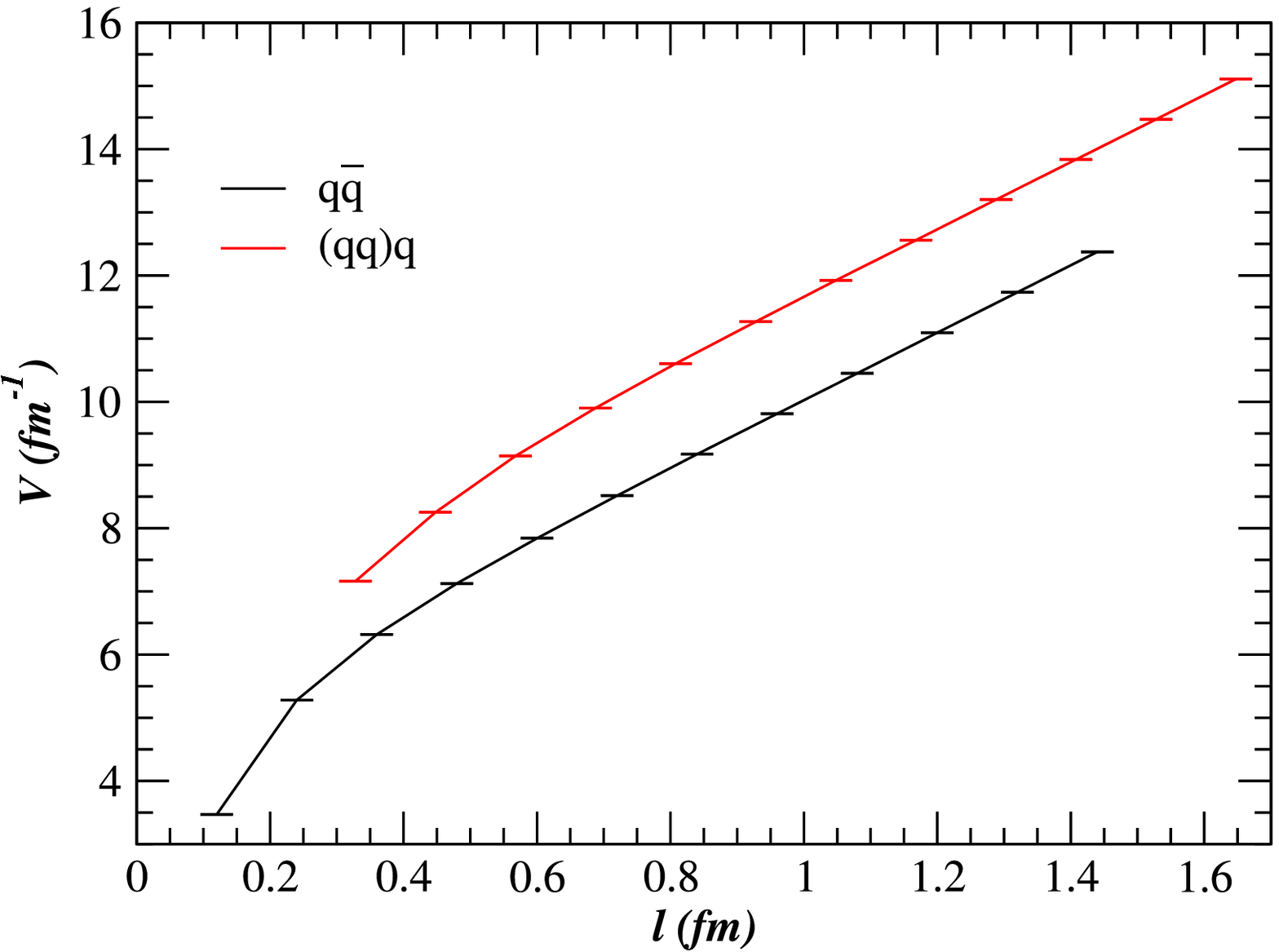}
\centering\includegraphics[width=4cm,clip=true]{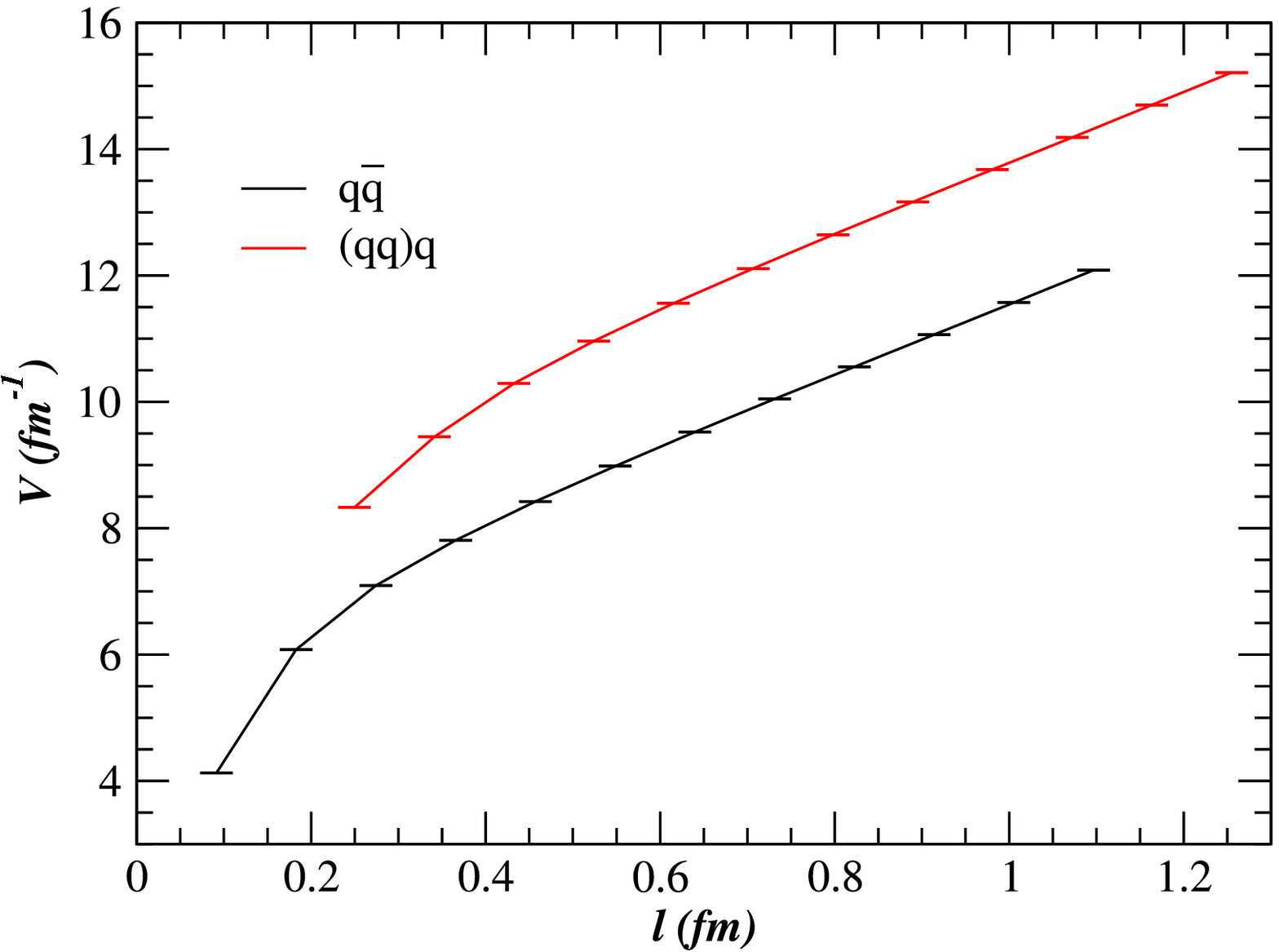}
\caption{Comparison of quark - antiquark and quark - diquark effective potentials for 
$\beta = 4.6$ (left) and $\beta = 4.8$ (right).}
\label{effpots}
\end{figure}

The effective potential is obtained from the Wilson loops in the standard manner: 
\begin{equation}
a\, V(\vec{r},\tau)= \ln\left(
\frac{W(\vec{r},\tau)}{W(\vec{r},\tau+1)}\right).
\label{pot}
\end{equation}
We obtain stable plateaus for the potentials as a function of $\tau$. 
%The statistical uncertainties are estimated using the jackknife method\cite{Montvay:1994bk}. 
The quark - antiquark potential has the well-known form
\begin{equation}
V_{Q\bar{Q}}(r) = V_{0} - \frac{\alpha}{r} + \sigma_{Q\bar{Q}} r
\label{potform2}
\end{equation}
where $\sigma$ is the string tension.
The three quark potential is \cite{Takahashi:2002bw,Alexandrou:2001ip}
\begin{equation}
V_{3Q}(r) = \frac{3}{2}V_0 -\frac{1}{2}\sum_{j<k} \frac{g^2 C_F}{4\pi r_{jk}}
+\sigma_{3Q} L(r) \, ,
\label{potform3}
\end{equation}
where $C_F =4/3$ and $L(r)$ is a length linking the quarks.  
As shown in our earlier work\cite{Bissey:2007tq}, $L(r)$ is given by the minimum length of string 
that connects the three quarks, or the sum of distances from the quarks to the Fermat (or Steiner) 
point.
%QCD predicts 
We expect that the two string tensions $\sigma_{Q\bar{Q}}$ and $\sigma_{3Q}$ are equal.
In Fig.~\ref{effpots} we plot the extracted effective potentials for the quark - diquark and 
quark - antiquark flux-tubes at each of the values of $\beta$ for our gauge configurations. 
The plots show that 
%the QCD prediction 
our expectation is confirmed at both values of $\beta$. 
Converting length measurements from lattice units to fermi we obtain the quark - diquark string 
tension $\sigma_{3Q} = 0.98 \pm 0.01$ GeV fm$^{-1}$, which is in excellent agreement with the quark - antiquark string tension $\sigma_{Q\bar{Q}} = 0.97$ GeV fm$^{-1}$.

\begin{figure}
\centering\includegraphics[width=4cm,clip=true]{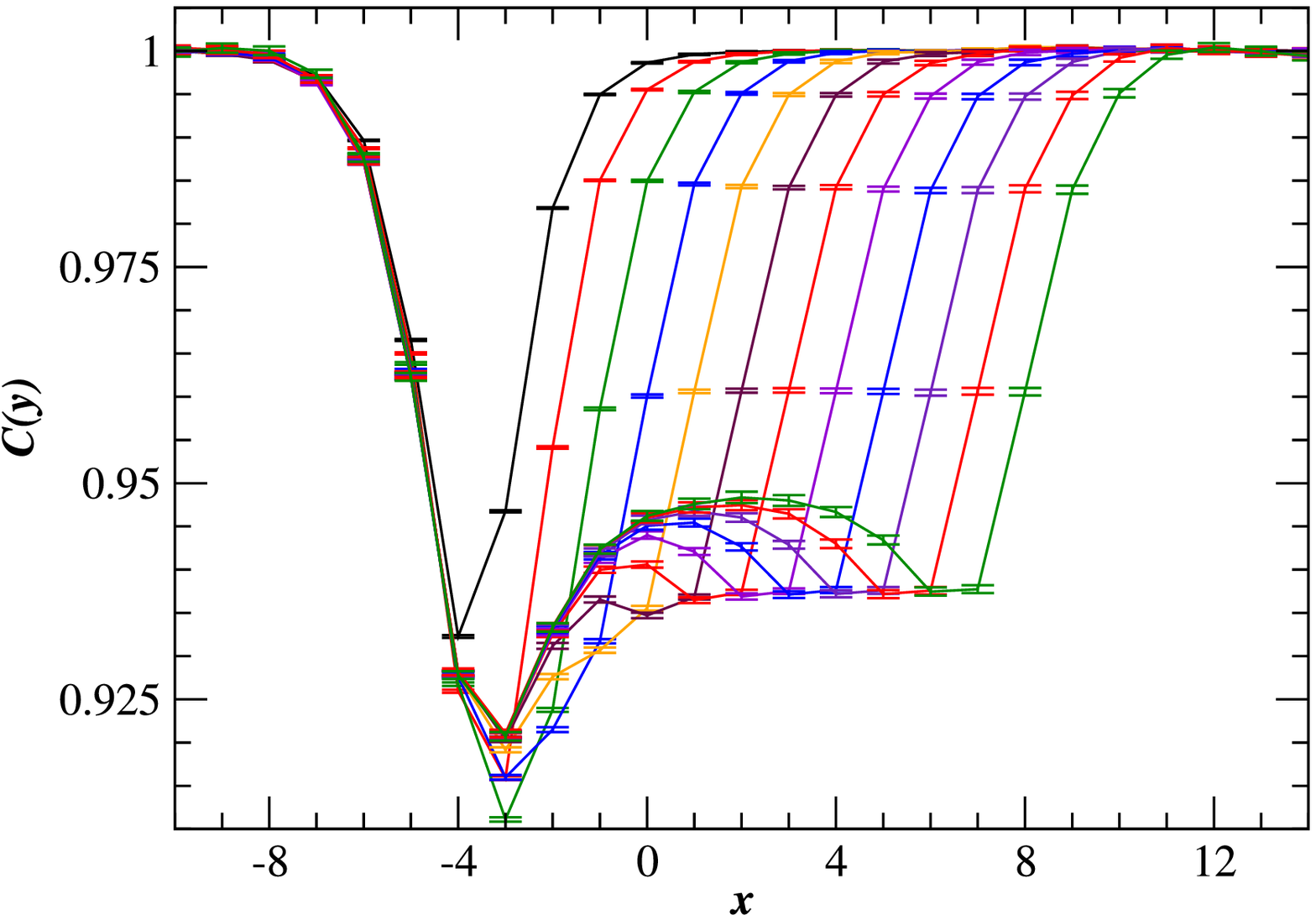}
\centering\includegraphics[width=4cm,clip=true]{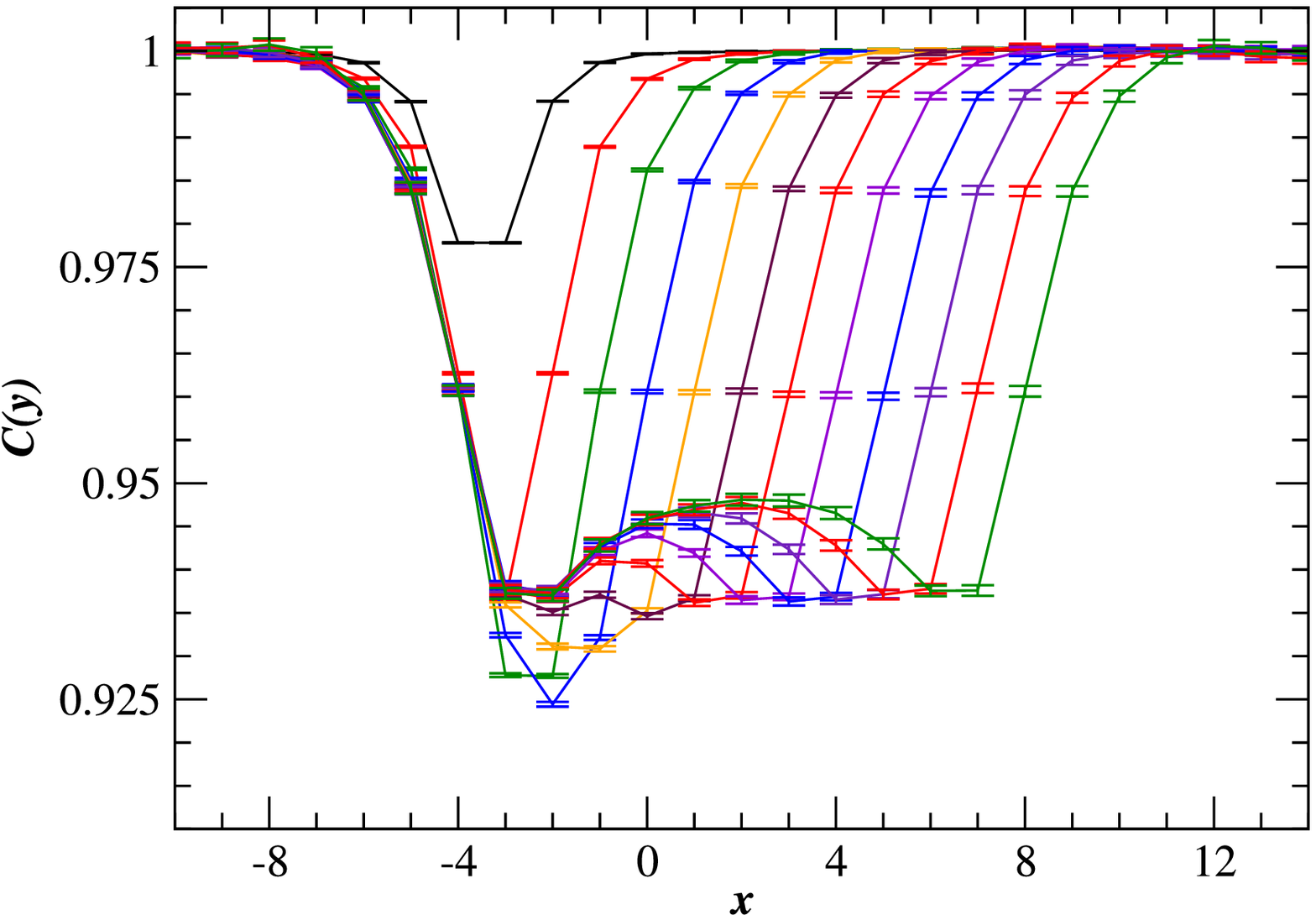}
\caption{Comparison of longitudinal flux tube profiles for quark - diquark (left) and 
quark - antiquark (right) flux tubes at $\beta = 4.6$ at all longitudinal separations.}
\label{longprofiles}
\end{figure}

We can gain further insight into the properties of the flux tubes by examining 
their profiles close to the quark.
We study the values of the correlators $C_{3Q}(\vec{y})$ and $C_{Q\bar{Q}}(\vec{y})$ where
$(\vec{y}) = (y_{1}, y_{2}, 0)$ is constrained to the plane of the color sources, and the origin is at 
the position of either the antiquark or the join of the T. 
The quark is then at the position $(\xi, 0, 0)$ where $\xi$ varies from 2 to 12 lattice steps.
First we examine the longitudinal profiles of both quark - diquark and quark - antiquark 
flux-tubes along the line $(\vec{y}) = (x, 0, 0)$ in Fig.~\ref{longprofiles}. 
As expected, the vacuum expulsion close to the diquark is stronger than in the vicinity of the 
antiquark. 
However, near the quark the two flux tubes show very similar profiles. 
Similar results are seen at $\beta = 4.6$.

\begin{figure}
\centering\includegraphics[width=5cm,clip=true]{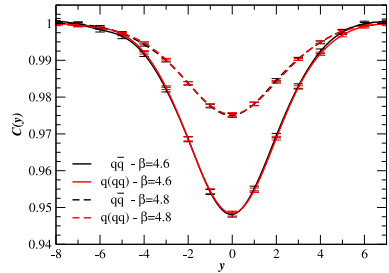}
\caption{Transverse profiles for quark - diquark (red lines) and quark - antiquark (black lines) 
flux tubes at $\beta = 4.6$ (solid lines) and $\beta = 4.8$ (dashed lines).}
\label{trprofiles}
\end{figure}

Next we examine the transverse profiles along a line orthogonal to the midpoint of the flux tube, 
{\em ie.} along $(\xi /2, y, 0)$ for $\xi$ even, or along  $((\xi +1) /2, y, 0)$ for $\xi$ odd.
In Fig.~\ref{trprofiles} we show profiles of both quark - diquark and quark - antiquark flux-tubes 
for $\xi = 12$. 
We find that as long as $\xi$ is larger than 4 lattice steps, the transverse profiles are close to 
identical. 
In earlier work \cite{Bissey:2007tq} we saw that the profiles saturate for long enough Euclidean time evolution.
Using a fit to a Gaussian profile we find that the transverse profiles of quark - diquark and 
quark - antiquark flux-tubes are statistically identical\cite{Bissey08}. 

\section{Conclusions}

We have directly compared gluon flux-tubes for quark plus antiquark and three quark systems. 
In the three quark systems we kept two quarks close together (two lattice units separation),
so that the system approximates a quark - diquark string. 
We found that the string tension in the quark - diquark string was the same as for the 
quark  - antiquark string.
In addition we compared the vacuum expulsion in both sets of flux-tubes. 
We found that, in the vicinity of the quark, there was no measurable difference between the  
transverse profiles of the quark - diquark flux-tubes and the quark - antiquark flux-tubes. 

These findings confirm the expectation from QCD that a diquark has many properties in common 
with an antiquark.
In particular the long range color interaction between a diquark and a quark is seen to be the 
same as that between an antiquark and a quark.
%This result is interesting in that it has been obtained in the quenched approximation, where the color 
%hyperfine interaction should be small. 
%This implies that the APE smearing and propagation in Euclidean time we have performed have 
%been sufficient for decuplet baryon states to decay. 
It would be interesting to repeat this work with dynamical quarks, where 
%variation in the strength 
%of the color hyperfine interaction could be investigated.
%This is potentially of great importance to phenomenological models of hadron structure where the 
%color hyperfine interaction is believed to be responsible for many observed properties of hadrons, 
%including mass splittings and magnetic moments.
the effect of screening by quark - antiquark pairs could be observed.

\section*{Acknowledgements}

This work has been done using the Helix supercomputer on the Albany
campus of Massey University and supercomputing resources from the
Australian Partnership for Advanced Computing (APAC) and the South
Australian Partnership for Advanced Computing (SAPAC).  The 3-D
realisations have been rendered using OpenDX
(http://www.opendx.org). The 2D plots and curve fitting have been done
using Grace (http://plasma-gate.weizmann.ac.il/Grace/).

\end{document}